# Home Advantage in the Brazilian Elite Football:
## Verifying managers capacity to outperform their disadvantage

By Carlos Denner dos SANTOS e Jessica ALVES

**Abstract**

Home advantage (HA) in football, soccer is well documented in the literature; however, the explanation for such phenomenon is yet to be completed, as this paper demonstrates that it is possible to overcome such disadvantage through managerial planning and intervention (tactics), an effect so far absent in the literature. To accomplish that, this study develops an integrative theoretical model of team performance to explain HA based on prior literature, pushing its limits to unfold the manager role in such a persistent pattern of performance in soccer. Data on one decade of the national tournament of Brazil was obtained from public sources, including information about matches and coaches of all 12 teams who played these competitions. Our conceptual modeling allows an empirical analysis of HA and performance that covers the effects of tactics, presence of supporters in matches and team fatigue via logistic regression. The results confirm the HA in the elite division of Brazilian soccer across all levels of comparative technical quality, a new variable introduced to control for a potential technical gap between teams (something that would turn the managerial influence null). Further analysis provides evidence that highlights managerial capacity to block the HA effect above and beyond the influences of fatigue (distance traveled) and density of people in the matches. This is the case of coaches Abel Braga, Marcelo Fernandes and Roger Machado, who were capable to reverse HA when playing against teams of similar quality. Overall, the home advantage diminishes as the comparative quality increases but disappears only when the two teams are of extremely different technical quality.

**Keywords:** Soccer. Team performance. Home advantage. Brazil. Football. Management.

# 1 Introduction

Commentators and fans readily assume there is a home advantage in soccer, a phenomenon documented in the literature of various sports and widely publicized in the media. Everyone expects the home team to win (and it does most of the time!). The explanation for such fact, however, is not at all clear. Ask commentators and fans for why this happens and watch a debate unfold; look for a precise answer in the literature and find no integrative explanatory theory. Is it because of the crowd presence, or the fatigue from traveling? Is there an advantage for the home team, or the away-team carries a disadvantage? Does the advantage disappear when the home-team is of lower technical level? Although much attention has been paid to answer these questions, few studies have indeed proposed an alternative to measure or study the advantages of playing at home. So, this paper develops a theory and a new model to answer these questions, explaining the home advantage with an analysis that includes match- and team-level data from 2003 to 2015, controlling for more than 20 teams and match facts simultaneously. More specifically, we analyze a panel data set containing unique information of thirteen years of match data from the website Soccer Way. The data were used to analyze the facts that contribute to HA of the highest performing teams in Brazil over this period.

Thus, these facts are related to five broad factors of performance in sports, namely, technical quality (past performance), fatigue (traveling distance), coach tactics (personal style), crowd presence (of rivals included), and referee behavior (cards and fouls). The investigation of factors influencing performance in any sport is crucial for its competitive evolution. Obtaining accurate and precise information about players and teams is in the interest of sports coaches and scholars, because it allows to relate the performance of and teams with the strategies and tactics used and, thus, contribute to the improvement of the training programs and the management of teams and players (Corrêa et al., 2002; Silva, 2012).

Professional soccer clubs are traditional labor organizations that deal with increasingly larger amounts of money. However, despite being managed by entrepreneurs and professionals, most of them demonstrate disappointing financial results (Silva, 2012). In this article, we aim to contribute to the ever-growing economic literature on the determinants of soccer home advantage by exploring the increasingly important yet underresearched data and proposing a novel model that explains the HA phenomenon.

The verification of this new integrative theory/model was based on the application of statistics and logistic regression to a unique dataset, assembled specifically for the model test. The analysis of players' behaviors in different contexts is one of the fundamental means for understanding the evolution of a sport (Garganta, 1994). From the analysis of relevant information about the game, what is sought is to optimize the behavior of players and teams in the competition as well as analyze the opponent's previous performance by knowing their strengths and weaknesses.

This knowledge, when systematized, allows to set up theoretical models, which enable not only to build more effective training methods but also point out evolutionary trends (Garganta, 1997). The main advantage of the chosen method over previously used methods is that our integrative theory did yield favorable empirical support. Since we treat HA as a decision-making situation, it is possible to incorporate the insights from the general decision making literature in a further attempt to build an integrative theory, model or framework of HA choice. Previously used methods have the effect of "regressing" each team's HA towards the mean HA for all teams combined, and therefore reduce the power to detect differences between teams. However, what is required is a maturity to outlook the HA as a complex situation and this paper proposes a new integrative theory, model and research to do that, based on five influential factors for soccer times.

The outcomes of that measurement model build a logically and consistent theory, with relatively superior explanatory power. Our results confirm the home advantage in the elite division of Brazilian soccer across all levels of comparative technical quality. The advantage diminishes as the comparative quality increases but disappears only when the two teams are of extremely different technical quality, something unlikely to occur in the same division. All broad hypothesized factors appear to play a relevant role in explaining the phenomenon, but each combination of factors specifically depends on the comparative technical quality of the teams, as the proposed theory states. There is evidence that at least one soccer coach was able to act and overcome the disadvantage of playing away from home regardless of comparative technical quality. Fans also influence by making the stadium dense, but only when the comparative technical quality between the teams is different. These findings provide ground to inform managers, coaches, players and fans about what to do to help their team perform better.

As contributions to the literature, the research approach here developed: 1) improves substantially the measure and calculation of home advantage previously adopted in the area, providing a more accurate alternative for future reference; and 2) provides an original-integrative theoretical model of team performance based on a so-far dispersed literature.

Thus, the present study aims to analyze the influencing factors on teams' performance in soccer, comparing it at home and away. The study considered five factors that seem to relate to the teams performance, classified in physical, technical, tactical and other match-related aspects, thereby explaining the possible home advantage (HA) of the host team, a well-documented fact in many individual and team sports (Courneya & Carron, 1992; Nevill & Holder, 1999; Pollard & Pollard, 2005; Jonhston, 2008).

## 2. Theoretical framework: Teams performance in soccer

The performance analysis is an area that has been consolidated for many years in the European soccer, with several studies analyzing team performance in European leagues (Boyko et al. 2007, Nevill et al., 1996; Page & Page, 2007). In Brazil, there are still a few clubs that have a structured department to develop this work (Montano, 2014) and few research. It is interesting for coaches and researchers to identify actions related to the effectiveness of teams and put them into practice (Garganta, 1997). Currently, the use of performance indicators assessment, defined by Hughes & Bartlett (2002) as a combination of variables that define aspects of performance, is used for both individual player and team analysis as for comparisons to opposing teams.

The tactical, technical and cognitive capabilities underlying decision-making are considered essential requirements for the excellence of sporting performance. Corrêa et al. (2002) identified in its research the contextual, psychological, technical, physical and tactical factors as those considered most important by respondents to the performance of the players. Among the physical aspects, they emphasized aspects related to training, physical preparedness and feeding of the players. The technical/tactical aspects correspond to the assimilation of the trainer's (coach) work method and improving these technical and tactical foundations. The elements identified by the authors indicate that learning processes are variables that assist in the constitution of the individual and group *performance* of athletes. Also, psychological factors such as confidence, motivation, and mental preparation; and social factors as the unity of the group, family, and organizational context are also associated with the performance of soccer players. According to Orlick (1986), athletes need three basic characteristics for excellence in sport: talent, intense training, and "head."

Thus, sports performance results from the combination of three factors: physiological, biomechanical and psychological. The athlete must be prepared physically, technically, tactically and psychologically to excel in his or her sports modality, which requires a planned work aiming at improving the requirements for obtaining better results (De Rose Junior et al., 2001). This is then the ***central hypothesis:*** *physical, technical, tactical and psychological aspects influences team performance.*

## 1.1 *The advantages of playing at home*

Several authors apply the concept of advantage of playing at home exclusively for soccer and suggest possibilities that provide this advantage to teams "at home" with the crowd's support, familiarity with the field, the favoring of referees, the displacement of visitors teams, the different tactics adopted by teams, psychological factors and even higher levels of testosterone in the players "at home" for territorial defense (Courneya; Carron, 1992; Nevill & Holder, 1999; Pollard, 1986; Brown et al., 2002; Clarke & Norman, 1995; Neave & Wolfson, 2003). Since the classic Schwartz and Barsky study in 1977, several theories have emerged as an attempt to explain which potential mechanisms may be conferring on home teams such an advantage. After studies have shown the behavior of HA in different soccer championships as the two main divisions of Brazilian soccer (de Almeida et al., 2011,) the two major continental championships (Drummond et al., 2014), and nationals in the world (Pollard and Gómez, 2009; Pollard and Gómez, 2014).

In England, for example, HA behaves differently in soccer depending on the team and division. Pollard & Pollard (2005) found values between 60% and 65%, with non-significant differences between the first and the second division in European countries. By analyzing Scottish and English leagues, Nevill et al. (1996) confirmed home advantage in the eight major divisions, though varying considerably across the divisions.

In Brazil, Pollard et al. (2008) pointed out that the found values (~65%) are higher than those from the main national leagues of Europe. Therefore, it is possible to analyze a competition to try to understand its competitive structure and tactical behaviors according to opponents who historically have an advantage at home (Drummond et al., 2014).

### 1.1.1 *Comparative Technical quality*

Even though it is important to consider effects in the game site to the identification of HA in the performance evaluation, some studies (Barnett and Hilditch, 1993; Clarke and Norman, 1995; Pollard and Gómez, 2009) have emphasized the necessity of adjusting the team capacity when quantifying the magnitude of the advantage at home for teams in sports such as soccer. For example, a problem arises when the HA calculation is based solely on match results or acquired points. If a strong team plays against a weak team, the difference in teams technical quality compensates for the relatively small effect of HA influencing the result (Gómez et al., 2013).

That is, the greater the difference in the qualities of the teams facing each other, the greater the probability that the stronger team will win both at home and out, masking the effect of the home advantage (Gómez et al., 2013; Hughes & Frank, 2005). Thus, when considering the effect of home advantage, the comparative/relative technical quality (RTQ) of teams should be taken into consideration.

**Hypothesis 1:** *The technical quality of teams influences the performance of the team at home.*

### 1.1.2 *Physical fatigue*

According to Fernandes (1994), currently, soccer demands a lot of the physical and mental capacities of players. It is common for talented players not to be able to show their abilities to the fullest due to bad physical conditioning. Therefore, physical preparation at soccer clubs is indispensable to the success of the team and for the performance of their players. Thus, the success of the development and maintenance of income in matches and competitions is due to training as well. Training directs the athlete's ability, leading him toward improving physical, technical, tactical, and psychological conditions (Fernandes, 1994).

Tied to physical training, resting is crucial for athletes to achieve superior performance. Authors have suggested methods and techniques for physical and psychological recovery of training and matches (Kellmann & Kallus, 2001). The proper use of recovery techniques accelerates regeneration and body restoration, which decreases the level of fatigue and the frequency of injury (Greig & Johnson, 2007; Moraska, 2007).

In 2014, the 8th Court of Work of Campinas condemned the Brazilian Football Confederation (CBF) for disrespecting the minimum period of 72 hours between matches of the same team. The decision was taken due to the lawsuit filed by the National Federation of Professional Football Athletes and aimed at protecting the health of players, enabling their muscle recovery between matches and avoiding injuries (Rodas, 2014).

Physical fatigue of the players is also related to the long and tiring trips made to certain games (Silva & Moreira, 2008). Pollard (1986) proposes that home advantage can be influenced by the distance traveled by the visiting team, which he called "*travel effect*". This effect has been investigated by some authors (Pollard, 1986; Clarke & Norman, 1995; Brown et al., 2002; Pollard, 2006; Pollard et al., 2008) in several countries, however, with contradictory conclusions.

Clarke & Norman (1995) showed that home advantage increased according to the traveled distance. The high value of HA found in the European Cup and in the Champions League (Pollard et al., 2005) and in competitions as the World Cup (Brown et al., 2002), can be consequence of long and tiring trips which can hinder physical yield of athletes, interfering with the performance of the team. However, Pollard (1986) showed that there is no difference of HA among teams that are less than 320 kilometers away from each other. Moreover, Pollard et al. (2005) verified a decline of HA found in various divisions of the English Championship, since trips have become easier and more comfortable over the years.

Silva & Moreira (2008) reviewed this factor in the Brazilian perspective and suggested that the high value of HA found for the Brazilian teams is related to the long distances traveled by the times of Brazil, which is a country with a large size and with great climatic differences among its regions.

The Pollard and Gomez (2007) method of estimating HA for individual teams was used in studies of Brazilian and Greek football. In the First Division of the Brazilian football league, significant variation in HA between teams was observed. In particular, teams in the north and south of Brazil had significantly higher HA than those from the central region; effects of travel and change in climate were suggested as possible explanations.

***Hypothesis 2:*** *Physical fatigue of players influences the performance of the team at home depending on the RTQ.*

1.1.3 *Coach's tactics and personal style*

The trainer or manager plays a significant role. He or she is a leader, motivator, facilitator and responsible for the decision-making. The coach has the final responsibility to mold the team. It is up to him or her to identify and cultivate the strengths and individual skills by promoting an environment in which athletes are motivated and productive in their roles (Half, 2014). Riemer and Chelladurai (1998) highlight the importance of how a coach utilizes the technical and tactical skill of players, selects and applies appropriate command strategies, trains and instructs athletes, and works individually with each player toward team performance.

Tactical principles are defined as a set of rules for the game that afford players with the possibility of developing solutions to the problems of the situation that confronts (Garganta & Pinto, 1994). For owning this character, tactical principles need to be understood and present in the behaviors of players during a match, so that their application facilitates attaining collective goals (Costa et al., 2009).

Corrêa et al. (2002) identified in interviews with professional players and coaches aspects related to actions that the Technical Commission should execute in order to provide favorable conditions for a good performance of athletes. Through these answers, factors that range from the tactical organization of the team, through emotional preparedness and group administration were raised, reaching the working methods themselves. In this sense, athletes have cited the importance of having a well-prepared, experienced technical committee that carries out plans, having an inclusive higher education. Moreover, respondents also commented on the importance of the trainer to make athletes assimilate their working methods, improving technical and tactical skills, maintaining discipline, harmonizing team sectors, organizing the team tactically and training all equally, both holders and reserves (Corrêa et al., 2002).

Pollard (2008) states that the strategies and tactics adopted by coaches are influenced by the place where the match occurs. Thus, coaches tend to adopt more offensive strategies when they play at home than when they play away (Dennis & Carron, 1999; Pollard et al., 2005). In this way, when playing at home, more action related to offense and assertion, such as shots are performed by the player of the local team (Mcguire et al., 1992; Varca, 1980). Thus, trainer's tactics and personal style can partially explain HA.

**Hypothesis 3:** *The coach tactics and personal style influences the team's performance at home depending on the RTQ.*

### 1.1.4 *Fans and crowds presence*

The support offered in stadiums psychologically influences both the performance of players and referee decisions, thereby contributing to home advantage (Nevill et al., 1999). Successful teams mention the great number of fans in stadiums as a positive influence on their performance. Players motivate the vibration of fans and value the opportunity to show fans their abilities (Gould et al., 1999).

The fans agree with that statement. Wolfson et al. (2005) found that the fans of the clubs of the English football leagues believe that home advantage is more influenced by crowd support than by familiarity, travel, territoriality, and referee bias. Furthermore, they associate the victory of the team with their support, as well as the distraction of opponents and influence on referee's decision.

Pollard (2008) suggests that fans effect is measured by its density (the presence of the fans expressed as a percentage of the total stadium capacity). Legaz-Giddyap et al. (2012) highlight that size, intensity, and proximity of fans with the field are factors capable of affecting the attention and psychological of players, coaches and referees, thereby affecting the performance of the team and partially explaining the effect of home advantage.

For instance, researches have demonstrated an association between crowd size and HA (e.g. Ponzo & Scoppa, 2014), while other studies have shown that it exists even with tiny crowds (e.g. Staufenbiel et al., 2016) and in one example with no crowd at all (Van de Ven, 2011).

***Hypothesis 4:*** *Fans and crowds influence the performance of the team at home depending on the RTQ.*

*1.1.5 Referees behavior*

Referees are designated to be impartial and fair, perceptive and intuitive, by partially out his or her emotions from the facts and situations (Half, 2014). However, Dohmen & Sauermann, 2016 found that local fans, through pressure and intimidation by the "homemade" environment, can psychologically exert influence on the referee's decisions in important actions of the game, in favor of the house team. Thus, Nevill et al. (2002) noted that, in the English league, referees who watched the video of the match with the noise of the crowd, marked fewer fouls for the home team, compared to referees who watched without noise influence. Other studies found that referees have indicated more penalty shootouts in favor of the commanding team and penalized visitors with more cards in this same league (Nevill et al., 1996; Boyko et al., 2007).

Furthermore, although the crowd may influence the performance of a team by putting home players in a more positive and confident psychological state, it has now been shown consistently that it also affects the performance of the referee who often is called upon to make game-changing decisions such as issuing a red card or awarding a penalty (e.g. Goumas, 2014b; Pollard &Armatas, 2017; Seckin & Pollard, 2008; Unkelbach & Memmert, 2010).

Johnston (2008), in turn, did not found a similar result. By replicating Boyko and colleagues' analysis in the English Premier Season at a different time frame he found no influence of referee's behavior on home advantage. Despite this negative result, which we believe that was because the author did not consider the relative technical quality effect, we suggest that:

***Hypothesis 5:*** *Referees behavior influences the performance of the team at home depending on the RTQ.*

The integrative theory, comprising all hypotheses developed in this paper appear in Figure 1.

**Figure 1**

Theoretical model: Factors influencing the performance of soccer teams.

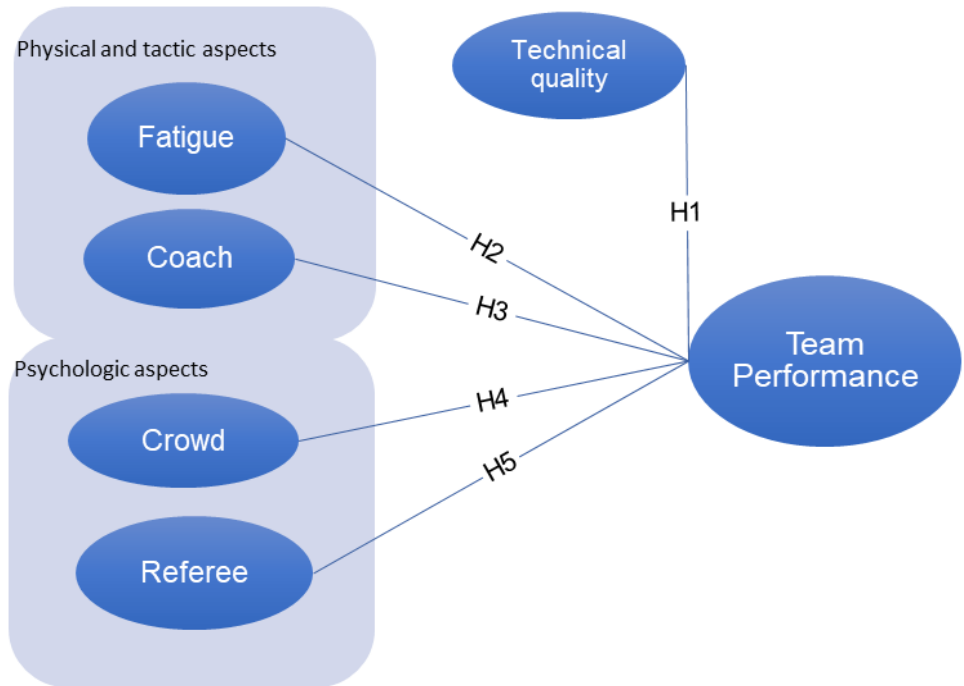

Source: The authors.

## 3. Research methods: General description and measurement

The present study is an explanatory research with quantitative approach. We analyzed teams that played in the elite division in the years 2003 to 2015 of the Brazilian Championship. These are twelve teams in total: Corinthians (SP), Flamengo (RJ), Cruzeiro (MG), Santos (SP), São Paulo (SP), Fluminense (RJ), Coritiba (PR), Atlético-PR (PR), Atlético-MG (MG), Internacional (RS), Grêmio (RS) and Goiás (GO).

The data were obtained via the website Soccer Way[1], which has been used in several studies for performance analysis and calculation of HA in soccer (Pollard, 2006; Pollard et al., 2008; Silva et al., 2008; Silva & Silva, 2010; de Almeida et al., 2011). The website has files with information about the world's main soccer leagues. For this study, information was consulted from the main Brazilian National League, the Brazilian Championship. The data was collected from September 16th, 2016 until October 11th, 2016. 1368 games are part of this sample in terms of victory, tie, and defeat, in which each team had 114 games analyzed.

---
[1] http://www.soccerway.com

The following information was collected from the site in each match for each team: data; round; number of goals in favor and against; Name, city and total capacity of the stadium; Quantity of players in defense zone, midfield and attack; number of red and yellow cards applied; and the name of the coaches. After the organization of the data, the analysis was done through the *software* SPSS.

In order to better understand the factors influencing the performance of a team at home and away, it was defined as a home match those held at the stadium of the team in the analysis. Likewise, it is said as a match out of the house those held at the adversary team stadium and the neutral field that game which was held in any stadium that does not correspond to any team faced before. Therefore, this study was disregarded the matches on neutral stadiums.

We chose to quantify the home advantage in two ways and compare them, in order to better understand the magnitude of this phenomenon. We did this as the measure for HA from Pollard could not represent HA in all its facets (it considers ties). The first way, a contribution of ours to the literature as it considers performance in both types of occasions exclusively as victories, will be called "HA per Wins" and consists in the comparison between home wins and away from home wins, demonstrated in Equation 1:

$$HA = \frac{Wins\ at\ home}{Number\ of\ games\ at\ home} - \frac{Wins\ away\ from\ home}{Number\ of\ games\ away\ from\ home} x100 \qquad (1)$$

For positive values, the advantage at home is greater than outside, for negative values the advantage outside the home is greater and, for values equal to 0 there is no advantage.

The second way will be called "HA per points" and consists in the method proposed by Pollard (1986) which measures the advantage at home from points obtained at home on the grounds of the total points obtained in all the matches in percentage, as in Equation 2:

$$VC = \left(\frac{(vc\ x\ 3) + (ec\ x\ 1)}{(vc\ x\ 3) + (ec\ x\ 1) + (vf\ x\ 3) + (ef\ x\ 1)}\right) x100 \qquad (2)$$

Wherein *HA* represents the number of wins at home; *EC* the number of draws at home; *VF* the number of wins away; *EF* the number of draws outside the house. Ratios are relative to the amount of points obtained depending on the outcome of the match[2]. The value 50% indicates no home advantage, since the same number of points has been acquired both at home and out. Thus, the greater the value above 50%, the greater the advantage of playing at home, and for values less than or equal to 50% the home advantage does not exist.

According to the conceptual model (Figure 1) presented, the performance analysis should take into consideration the quality of the teams. The technical quality of each team was calculated with the aid of Equation 3:

$$Q = \frac{\sum_{2003}^{2012} Pc}{\sum_{2003}^{2012} Pd} \times 100 \qquad (3)$$

In which, *Pc* is the score conquered, and *Pd* is the score disputed by the team. Thus, quality is defined as the average yield of the team in the period from 2003 to 2012. This period was chosen, because, in previous years, the Brazilian Championship had a different dispute system, and until 2012, it did not consider the period of analysis. For this calculation was considered only the games of the Series A. Thus, if a team was demoted, it was disregarded its quality in the year(s) of demotion. The relative technical quality was calculated by the ratio between the quality of the team and its opponent. It was considered teams of the same quality: those that ration was in the range of 0.9 to 1.1. Below 0.9 was considered lower quality team and above 1.1 top quality time. The dependent variable "performance of teams at home" is represented by the variable "home wins" which assumes value 1 when it occurs and value 0 when it does not occur. To quantify the physical fatigue of the players, the variable "fatigue" was created. It is calculated through the division between the distance traveled by the team to a match outside and the number of days that the team remained without playing between one match and another. To evaluate the coach tactics was conducted an analysis of the matches by coach, considering each level of quality, to identify the HA of each and correlate them with the dependent variable. Thus, the name of the trainer when significant represents that this has an influence on the performance of the team.

---

[2] By the Brazilian Championship regulation, teams receive three points for victory and one point per draw. No points for defeats are attributed.

The stadium density was calculated through the division between the public present at the stadium in a given match and the total capacity of the stadium. The fans of the team and the opponent represent the percentage of the total people who support a given team, according to the survey conducted by Data Folha[3]. The referee behavior was also evaluated by three variables. The first was "red card balance" that was calculated by subtraction between the red cards applied to the team and the red cards applied to the opponent. The negative value means that the judge punished more the opposing team. The variable "fouls" refers to the amount of fouls committed by the team and the variable "adv fouls" the amount of fouls committed by the opposing team. To capture the moderating variable "technical quality" (H1), four analyses were conducted in different match situations: when the relative technical quality of teams is inferior, when it is equal, when it is superior, and ultimately when the quality is not taken into consideration.

The method used in this study was based on the application of statistics and logistic regression to a unique dataset, assembled specifically for the model test. This has the advantage over previously used methods because it allows the verification of a new integrative theory/model, and therefore maximizes the statistical power of the analysis, providing the highest performing decision for a time.

## 2 Results and discussion

The total research sample was composed of 1368 matches, in which a total of 260 matches in the neutral stadiums were verified (19%). It was observed that the occurrence of home wins (24.5%) is higher than the occurrence of outside wins (10.4%), as well as the occurrence of home losses (7.5%) is inferior to the occurrence of outside defeats (19.2%). Thus, through equation 1, a percentage difference of 33.5% to HA was found for victories. This positive value indicates that teams earn more often at home than outside. However, through Equation 2 was found the value of 66.5% for the HA per points, indicating that more than half of the points obtained by the teams in the championships were in matches at home.

These results show that, in general, the teams have the advantage of playing at home without taking the comparative technical quality into consideration. In matches with teams of same quality, the HA per win is 31% and 68% per points; in superior quality games, these numbers are 35% and 64%, and in lower quality games, 37% and 79%. Teams presented a higher advantage in home matches in which they had lower comparative quality, with both methods. In superior quality games, the HA per points was the smallest found (64.4%), and the HA per victories was the second minor (35.4%). Therefore, it is possible to indicate that teams presenting "weaker" features present a greater advantage of playing at home than when they are "stronger." In order to deeper understand the effect of home advantage, the HA analysis per team appears in Table 1 with the HA values, and quality found for each team in descending order.

---

[3] Source: http://www.rsssfbrasil.com/miscellaneous/torcidas.htm. Accessed on 28 sep. 2016.

**Table 1**

HA Rankings and technical quality.

| # | Ranking 1 | HA per victories | Ranking 2 | HA per points | Ranking 3 | Technical Quality |
|---|---|---|---|---|---|---|
| *1* | **International** | 54% | Coritiba | 75% | São Paulo | 57,75% |
| *2* | **Santos** | 49% | Santos | 74% | International | 52,42% |
| *3* | **Grêmio** | 47% | Atlético-MG | 68% | Cruzeiro | 51,83% |
| *4* | **Atlético-MG** | 42% | Fluminense | 68% | Corinthians | 51,47% |
| *5* | **Corinthians** | 35% | Grêmio | 67% | Santos | 51,42% |
| *6* | **Coritiba** | 31% | Corinthians | 66% | Fluminense | 49,58% |
| *7* | **Cruzeiro** | 31% | Goiás | 65% | Grêmio | 48,70% |
| *8* | **Atlético-PR** | 28% | São Paulo | 65% | Flamengo | 47,83% |
| *9* | **Flamengo** | 28% | International | 64% | Goiás | 46,50% |
| *10* | **Fluminense** | 24% | Flamengo | 64% | Atlético-PR | 46,41% |
| *11* | **São Paulo** | 21% | Cruzeiro | 63% | Atlético-MG | 45,40% |
| *12* | **Goiás** | 18% | Atlético-PR | 54% | Coritiba | 45,10% |

As HA values have always been positive in ranking 1 and above 50% in ranking 2, one can say that all teams have presented the advantage of playing at home in matches, at least with an analysis where relative quality has not been taken into consideration. It is interesting to note that Coritiba possesses the worst quality among teams and the greatest advantage at home per points. São Paulo and International, teams with greatest qualities, have not presented such great advantages compared to the other teams. It seems that the highest performing teams must be consistently good, in and out. In other words, the home advantage needs to disappear for a team to be successful over the years.

Taking into consideration the relative quality between the teams, figures 2 and 3 present HA indexes per wins and per points in matches with different technical qualities. The Santos team with 100% HA is the one that presented the highest advantage at home. Coritiba, the team with the worst quality, was the second in the ranking of HA per points, with high advantage (92%), and then Atletico-PR with 87.5% which also presented one of the lowest qualities and worst advantage when the quality was not considered. The value of 50% presented by Cruzeiro, the last of the HA ranking per points, shows that half of the points obtained were at home and the other half outside, proving that the team did not possess the advantage of playing at home in matches with inferior quality. The same is not observed by the HA of wins (16.6%), which indicates that the team has presented the advantage of playing at home. This difference occurred because Cruzeiro won the same amount of times at home and outside, even though it had the amount of outside matches higher than the amount of games at home, which caused this percentage difference. Grêmio presented HA for wins of 0% indicating the non-existence of advantage but presented HA per points of 57%. This is explained by the fact that the team has obtained points of the tie at home, which is counted through the point's method but not by the winning method. Both measures provide a more complete picture of the complexity of the home advantage phenomenon.

Atletico-PR was the only team that did not show an advantage at home playing in both methods of evaluation (-7.14% and 33.3%), presenting negative value of HA, which indicates a greater advantage of playing outside when possessed superior quality than of their opponents. Grêmio was the only one which did not present any kind of advantage (0%), and Santos was the only one presenting 100% of home advantage in some match situation. In conclusion, as for each type of confrontation, it was obtained a different ranking, with different values of HA found for each team, you can say that the technical quality of the teams interferes with the performance of the teams at home.

**Figure 2**

HA per team by the method of wins.

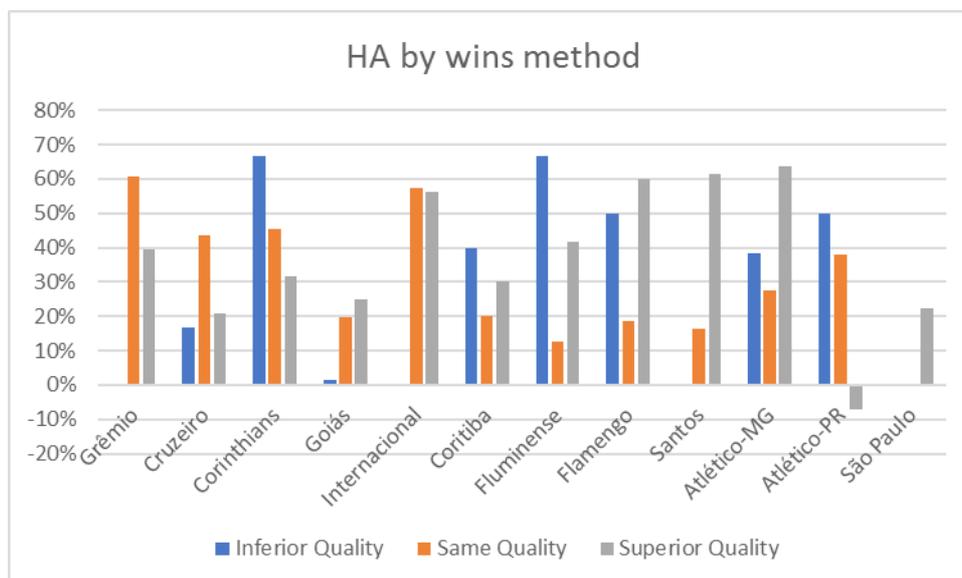

**Figure 3**

HA per team by the method of points.

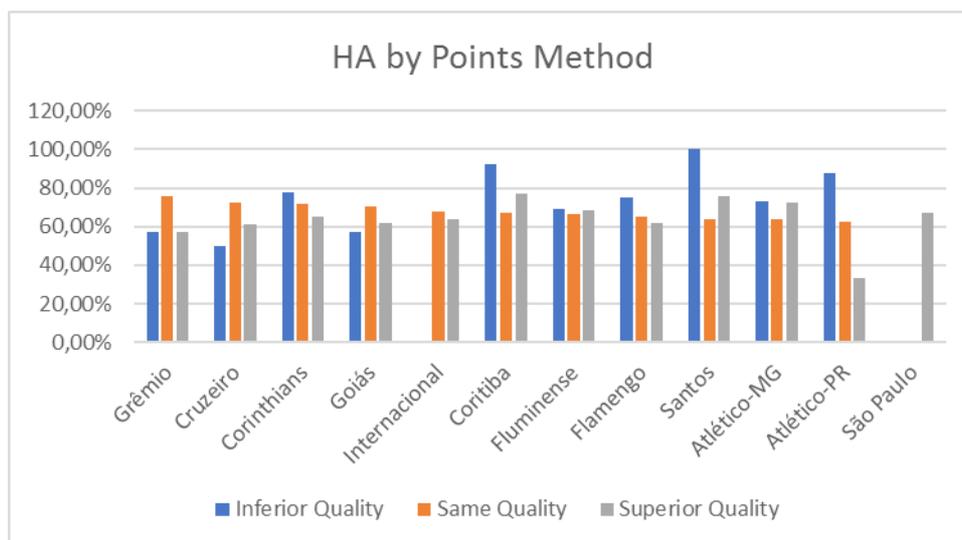

To test all the hypotheses and theoretical model, it was performed a logistic regression to verify the influence of all factors at the same time. Thus, the relationship of the dependent variable "performance" binary (yes or no) was verified with all the independent variables (factors). To check the effect of the moderator variable (technical quality), the test was performed four times for each relative technical quality case. Table 2 presents the results of the statistical tests conducted by highlighting only the coefficients found statistically significant. The table was divided into sections according to the RTQ. Only variables with a significant P-value were presented (P<0.10).

None of the independent variables showed significance in matches with lower relative technical quality, and so they are not presented in Table 2. This means that none of the factors accounted for in the analysis can explain the home advantage when the comparative technical is low. Not much can be done when the home team is too worse than the visiting team.

The models specified by the study managed to explain 46.6% of the performance variance of soccer teams at home when technical quality of the team is inferior; 41.2% when technical quality of the team is the same as the opponent; 46.5% when technical quality is superior; and, finally, 40.8% when technical quality of the teams is not taken into consideration.

**Table 2**

Logistic regression results.

|  | Independent variables | Coefficient (P < 0.10) |
|---|---|---|
| **Same quality** | Coach Abel Braga | 1.609 |
|  | Coach Marcelo Fernandes | 1.539 |
|  | Coach Roger Machado | 2.356 |
|  | Red Card Balance | 0.766 |
|  | Fouls | - 0,111 |
|  | Adv Fouls | 0.089 |
|  | N = 613; Cox & Snell R = 0.412; Nagelkerke R = 0.631 | |
| **Superior Quality** | Fatigue | - 0.059 |
|  | Coach Abel Braga | 1.89 |
|  | Coach Dorival Júnior | 3.03 |
|  | Public Density | - 1.654 |
|  | Adv Fans | - 16.633 |
|  | Fouls | -0.115 |
|  | Adv Fouls | 0.077 |

|  | N = 540; Cox & Snell R = 0.465; Nagelkerke R = 0.675 | |
| --- | --- | --- |
| **No Quality** | Fatigue | - 0.079 |
|  | Coach Abel Braga | 1.628 |
|  | Coach Argel Fucks | 2.352 |
|  | Coach Dorival Júnior | 1.863 |
|  | Coach Luiz Felipe Scolari | 1.112 |
|  | Coach Roger Machado | 1.671 |
|  | Public Density | - 0.967 |
|  | Adv Fans | - 6.088 |
|  | Red Card Balance | 0.43 |
|  | Fouls | - 0.1 |
|  | Adv Fouls | 0.073 |
|  | N = 1356; Cox & Snell R = 0.408; Nagelkerke R = 0.609 | |

It is possible to observe that the strength and magnitude that independent variables exert on the dependent changes according to each level of technical quality. The quantity of significant variables also changes depending on quality. For example, in games that technical quality is not taken into consideration, the quantity of significant independent variables is much greater than in other cases. As verified in Figures 2 and 3, there was a difference in the HA behavior of each team depending on the technical quality. It concludes that each team presents a certain advantage depending on its own quality and also the technical quality of its opponent, then confirming the hypothesis 1 of this study.

The variable "physical fatigue" is significant in higher quality matches and in matches where the quality is not considered, with a negative effect on the dependent variable. This shows that the little rest of the players and the long journeys traveled by the team interferes in a negatively in performance by diminishing the probability of the team winning even playing at home. This result confirms Hypothesis 2 (the physical fatigue of players influences the performance of the team at home depending on the relative technical quality).

Concerning to coaches, six showed a positive influence on the performance of the teams at home. The coach Argel Fucks for example, with a coefficient of 2.352, exerts a much greater positive effect than the other coaches in the analysis, in matches that the technical quality is disregarded. Coach Abel Braga has shown himself influential in his home performance in the three-game situations. Dorival Júnior and Roger Machado were also significant in at least two situations. Thus, these coaches seem to present some technique/tactic or personal style that somehow benefits teams in which they train when they play at home. The significant relationship of some coaches with the team's performance at home confirms Hypothesis 3 of this study (the coach's tactic and personal style influences the team's performance at home depending on the relative technical quality).

The fans and crowd presence did not show influence on the performance of the teams at home independent of the technical quality of the teams, which is different than expected. On the other hand, the public present at the stadium (public density) presented a negative influence on the performance of teams at home in matches with superior quality and without quality. A fact that drew attention was the strong negative effect that fans of the opponent team exerted on the performance of the house teams, especially in matches where the quality was superior (-16.633). Therefore, hypothesis 4 (fans and crowd presence influence the performance of the home teams depending on the technical quality) is confirmed.

Finally, relating to referee behavior, the variable "red card balance" has been identified as significant in an unexpected way. There was a positive coefficient of significance in the same quality games and no quality games. Thus, if the red card balance is negative, it means that the opponent was more "punished," then the effect that this "punishment" exerts on the performance of the house team is negative. This result shows that if the referee applies more red cards to a given time, it is more likely that this team will have superior performance at home (perhaps for being more aggressive than its opponent). Similarly, in case the opponent receives more red cards, the probability of the home team gains decreases. Unlike red cards, the fouls committed by the team exert a negative effect on the performance of the home team, while the adversary's fouls exert a positive effect. These results confirm hypothesis 5 of this study in which referee behavior influences the performance of home teams.

Therefore, even if none of the variables were significant in lower-quality games, they were significant in at least one of other situations. The aims of this study were two-fold. First, improves substantially the measure and calculation of home advantage previously adopted in the area, providing a more accurate alternative for future reference; and second, provides an original-integrative theoretical model of team performance based on a so-far dispersed literature. Thus, all the hypotheses proposed by this study were confirmed by the statistical test, and in conclusion, the integrative theoretical model here developed can be considered valid, becoming a useful resource for strategic decision-making.

# 3 Conclusions and Recommendations

This study aimed to identify the factors influencing the performance of soccer teams inside their home. In this sense, the conceptual model, integrating previous research in the area, was proposed admitting the central hypothesis that the performance of sports teams is influenced by various aspects: physical, technical, tactical and psychological. With that in mind, it was developed a model to explain team performance in soccer, considering five influential factors: the physical fatigue of athletes, the tactic and personal style used by the coach, the effect of the supporting fans and the referee behavior influence, always taking into consideration the comparative technical quality of teams. It was concluded that home advantage is indeed present in Brazilian soccer, confirmed both by a traditional measure of HA, and our own original calculation, which is the main contribution of the present study.

The significant difference found in home advantage among the clubs is noteworthy and agrees with previous literature on the phenomenon. For example, Clarke (2005) found that in the Australian soccer league, non-Victorian teams have large home advantages. Similarly, Nevill et al. (1995) found that home advantage varies across English and Scotish soccer divisions.

To take advantage of this knowledge sets forth in this paper, visitor teams should explore the adverse circumstances they face. The technical committee has to develop tactics and counteroffensive strategies, as there is the offensive pressure of the home team and the influence of the referee. In this way, the visitor team could adopt a tactic that favors dribbling, ball passes and goal defense. On the other side, for home teams, knowledge and exploration of this advantage become an extra tool to offer difficulties to their opponents. Thus, managers and coaches could adopt an offensive tactic and choose the players with a more offensive profile to play at home. Supporters, in turn, would be more motivated to attend games played away if they knew that their presence could be fundamental for the performance of their teams. This could be fostered by the clubs by means of incentives, such as souvenirs and meeting with the players, in these games.

Among physical aspects, it has been identified that the greater the physical fatigue of players, worse the performance of the team at home. It is then suggested that physical trainers and coaches should have the notion of the importance of the physical and psychic recovery of athletes as a factor that favors increased performance. Thus, teams could invest more in hiring professionals related to the well-being of the players, as well as avoid keeping the same players on the base team for a long time. Another practical action that may diminish the effect of physical fatigue is to arrive at the adversary's location many days before the match so that the players could rest adequately. Concerning the trips, they should be as less stressful as possible and provide maximum comfort to the players. (Wolfson and Neave, 2004).

Regarding the psychological aspects, the effect of fans behaved somewhat differently than expected. While the opposing team fans were very influential in the performance of the home team, the fans of the team itself seem to have no significant influence. It seems that the noise of the contrary crowd has a much higher power (and negative) in the performance of the team than the support that fans offer. Thus, in order to avoid being negatively influenced by hostile crowds, visitor teams should focus on mental preparation, concentration and discipline of their players (Wolfson and Neave, 2004).

Besides that, the "full house" factor exerted a weak influence on performance. This might happen because "full house" is observed mainly in matches against strong teams, which usually perform well away from home, as suggested by Jonhston (2008). Alternatively, it is important for home teams to find ways to neutralize the effect of the opposing team fans. A suggestion is to mobilize local fans to attend in huge numbers so that opposing fans would be the vast minority. In this way, the fans would be more motivated to attend games played at home if they knew that their presence could be fundamental to the performance of their teams. This could be fostered by the clubs by means of incentives, such as souvenirs and meeting with the players.

Referee behavior also showed influence on the performance of players at home. As fouls committed by the home team are related to its lower performance, coaches should alert players of this importance while adopting an offensive posture at home, of course avoiding violent behavior. They should also be attentive whether the opposing crowd usually makes plenty of noise which could influence the referee's decision making (Nevill et al, 2002). Future studies should investigate the effect of fans on the referee in order to identify the existence or not of a possible arbitral advantage for home teams according to specific conditions, they should also investigate whether the profile of the referee (e.g., age, years of experience, nationality, gender) could minimize or even nullify the effect of arbitral advantage.

In conclusion, it is noteworthy the great importance of understanding home advantage in soccer so that managers and coaches can adopt strategies that diminish the negative influence of opponents fans; create routines to generate a familiar and friendly environment; avoid fatigue associated with long and little rest trips; take into account referee behavior; and promote a positive psychological state-of-mind of players. Therefore, it is suggested to the managers and leaders of soccer in Brazil to orient themselves based on the factors cited in this study to manage their soccer teams and organizations. Improvements in the physical structure of the club, hiring a good medical team and physical trainers, are ways to improve the performance of players since physical fatigue influences performance. Moreover, accompanying the team progress, hiring good players and coaches, attracting sponsors, fans and associates are ways to achieve higher technical quality (through long-term performance), factor proved crucial to explain the home advantage in soccer. It is fundamental to professionalize management of clubs to boost profits and results for the club-company. In this sense, for future studies, the financial issues of clubs should be taken into consideration as performance influencer as well. This study was limited in time and place, presenting a relatively short period of analysis (2003 to 2015) of a single country (Brazil). For future studies, it is suggested the investigation of a longer period of various countries, thereby increasing the sample.

Another issue that could be studied is the influence of violent crowds on the performance of the team. In this way, it is relevant to analyze if this behavior could influence the referee's decision or the own behavior of the players thus affecting home advantage. From our results, we believe that violent behavior could influence negatively home advantage if expressed by the home team since it might results in punishments to the team and even cause physical damage to players and staff. There is also the limitation of the variable "physical fatigue", which includes only the distance traveled by the team, not taking into consideration the conditions of travel and lodging, for example. In this way, sometimes a smaller distance traveled can be much more tiring, for being carried out by buses for example, than those larger distances traveled by airplane. Moreover, the distance traveled is calculated only as the distance between the city of the visiting team and the city of the commanding team, not taking into consideration that often the visiting team does not return to their city, leaving directly for upcoming matches in other cities. It is suggested that in future studies, a complete itinerary of the teams is made so that this variable of fatigue is better measured. Additionally, when analyzing team performance, context and social aspects should be taken into consideration. Factors such as the group's union and relationships between players with the coach and the team with the fans and the media should be studied. It is also suggested that interviews with professionals from the area are used as another way of obtaining data.

Although this study has provided some new insight into an understanding of the effect of HA in soccer, it does contain some limitations. As mentioned earlier, HA is a phenomenon influenced by many factors, operating simultaneously, before and during a match. The analysis in this study was constrained by the limited amount of five influential factors: the physical fatigue of athletes, the tactic and personal style used by the coach, the effect of the supporting fans and the referee behavior influence. Future research could look more specifically for other factors, for example, the effect on home advantage when a team changes from grass to artificial turf as recently done by Diniz da Silva, Braga, & Pollard (2018). Since 2016, Atlético-PR has a synthetic grass stadium, and is currently the only high performance club in Brazil to use that. However, our study includes data from 2003-2015, that is why this factor was not included in integrative method.

Despite the importance of Brazilian soccer in the country's economic and cultural scenario, knowledge about factors involved in soccer confrontations and its consequence in the form of advantage for the commanders is very little (Silva et al., 2010). Thus, further studies are needed on HA in Brazilian soccer, including in the different divisions that make up the professional soccer sector.

**Conflict of interest statement**